\documentclass[twocolumn,showpacs,prl]{revtex4}
\usepackage{amsmath,amsthm,amscd}
\usepackage{dcolumn}

\usepackage{graphicx} \usepackage{epsfig}

\begin{document}

\title{Simple microscopic model of the magneto-electric effect in non-collinear magnets}
\author{T. A. Kaplan and S. D. Mahanti}
\affiliation{Department of Physics and Astronomy and Institute for Quantum Sciences, \\Michigan State
University, East Lansing, Michigan 48824}

\begin{abstract}
A 2-site 2-electron model with s- and p-states on each site, having constrained average site-spins
$\mathbf{S}_a,\mathbf{S}_b$, with angle $\Theta$ between them, simpler than the previous closely-related model
of Katsura et al (KNB), is considered. Intra-site Coulomb repulsion and inter-site spin-orbit coupling $V_{SO}$
are included. The ground state has an electric dipole moment $\pi$ consistent with the result of KNB,
$\pi\propto\mathbf{R}\times(\mathbf{S}_a\times\mathbf{S}_b)\equiv\mathbf{f}$, $\mathbf{R}$ connecting the two
sites, and application to a spiral leads to ferroelectric polarization with direction $\mathbf{f}$, also in
agreement with the previous result. However, the present result shows the expected behavior $\pi\rightarrow0$ as
$V_{SO}\rightarrow0$, unlike the previous one.
\end{abstract}


 \maketitle
\section{Introduction}
The magneto-electric (ME) effect, i.e. the production of an electric polarization (magnetic moment) by
application of a magnetic (electric) field, has been known for a long time. Multiferroics, e.g. materials that
show ferromagnetism and ferroelectricity coexisting, have also been known where spins and electric dipoles order
at different temperatures. Quite recently a new class of multiferroics has been found, in which the magnetism
and ferroelectricity are strongly coupled; e.g., they order at the same temperature. For this effect, however,
the magnetic structure cannot be simple ferromagnetism or a collinear spin state; the (vector) magnetization
density must show spatial variation in its direction as, e.g. for a spiral spin state. I refer to the
introductory paragraphs of several recent papers for a more detailed history and list of references.~\cite{
sergienko,mostovoy, katsura, tokura,yamasaki} Until very recently, the effect has been found in
antiferromagnets, e.g., simple spirals, with no net magnetic moment. In~\cite{yamasaki}, however, the magnetic
structure of the material studied, CoCr$_2$O$_4$, is approximately a ferrimagnetic spiral (there is a net
moment).~\cite{menyuk,lyons, tomiyasu, kaplan}

The work of Mostovoy~\cite{mostovoy} is a phenomenological theory (see also~\cite{lawes}), whereas microscopic
models exhibiting the effect are presented in~\cite{sergienko} and \cite{katsura}. Both of the latter involve
superexchange (where electron hopping between magnetic ions involves an intervening oxygen ion).
In~\cite{katsura} the situation is considered where a nearest neighbor pair of magnetic ions has an inversion
center at the mid-point, so that there is no Dzyaloshinsky-Moriya interaction
(DMI)~\cite{dzyaloshinskii,moriya}; the polarization comes from a distortion of electronic density without ionic
or atomic displacements; also the $t_{2g}$ orbitals considered on each magnetic site are chosen to diagonalize
the spin-orbit coupling (\emph{intra-atomic} SO coupling), partially removing degeneracy of these orbitals. In
contrast, the essential mechanism invoked in~\cite{sergienko} depends on the existence of the DMI plus
electron-lattice interactions and orbital degeneracy (Jahn-Teller effect), and the polarization results from
ionic displacements.

It seems sensible to ask, is it necessary to consider such complicated models to obtain an essential
understanding of this unusual ME effect? We present here a much simpler model which yields the effect, namely
the
 electric dipole moment for a pair of magnetic sites,~\cite{katsura}
\begin{equation}
\pi \propto \mathbf{R}\times(\mathbf{S}_a\times\mathbf{S}_b),\label{0}
\end{equation}
where $\mathbf{R}$ connects the sites a and b, and $\mathbf{S}_a,\mathbf{S}_b$ are the average spins at the
sites. It is most closely related to that of~\cite{katsura}, in that it considers the interaction of two
magnetic sites with a center of symmetry (so there is no DMI). Also, the source of the spatial variation of the
ordered spin density is due to outside effects of spin-spin exchange interactions, as in~\cite{katsura}. The
hopping is direct--there is no oxygen, and no ionic displacements or orbital degeneracies. It involves
\emph{inter}-atomic spin-orbit coupling, a mechanism different from the model of~\cite{katsura}; it is obviously
different from that of~\cite{sergienko}. We add that these characteristics are more appropriate than the others
for CoCr$_2$O$_4$, where there is no orbital degeneracy.\cite{lyons,kaplan}

A further motivation is  a difficulty with the work of~\cite{katsura}. In the 3d transition metal ions, the
spin-orbit coupling $V^{SO}$ is the smallest of the various energies, namely Coulomb interactions, transfer or
hopping integrals, and (cubic) crystal field splitting. But in~\cite{katsura}, $V^{SO}$ is taken as essentially
infinite: it is diagonalized, along with the crystal field, before the hopping energies are considered. This
leads to a doublet ($\Gamma_7$) and a higher energy quartet $(\Gamma_8)$, and the quartet is dropped. This is
probably the reason why $\pi$ does not show the expected vanishing when  $V^{SO}\rightarrow0$. Our model does
not have this difficulty.

Our simple model has two atoms or ions, a and b, lying on the x-axis; each has an s-type orbital and 3 p-type
orbitals lying an energy $\Delta$ higher. Also, there are two electrons, 1 per site. Later we describe briefly
the generalization to the case where each site has 3 $t_{2g}$ electrons, as for Cr$^{3+}$ on an octahedral site,
appropriate to the B-sites in XCr$_2$O$_4$, X=Co,Mn.

For our simple model we first constrain the 1-electron basis set (including spin) to make the average spins
$<\mathbf{S}_a>,<\mathbf{S}_b>$ lie at particular angles in the x-y plane and make an angle $\Theta$ between
them. The 1-electron basis states are
\begin{eqnarray}
s_a&=&u(\mathbf{r}-\mathbf{R}_a)\chi_a, \ \ s_b=u(\mathbf{r}-\mathbf{R}_b)\chi_b \nonumber\\
p_a^\nu&=&\nu v(\mathbf{r}-\mathbf{R}_a)\chi_a,\ p_b^\nu=\nu v(\mathbf{r}-\mathbf{R}_a)\chi_b,\label{1}
\end{eqnarray}
where the spin states are
\begin{eqnarray}
\chi_a&=&(\alpha+e^{i\phi_0}\beta)/\sqrt{2} \nonumber\\
\chi_b&=&(\alpha+e^{i\phi_b}\beta)/\sqrt{2}.\label{2}
\end{eqnarray}
($\alpha, \beta$ are the usual ``up, down" states along the z-direction.) And we take $\phi_b=\phi_0+\Theta$ as
in FIG. 1. This spin arrangement is chosen for simplicity.
\begin{figure}[h]
\centering\includegraphics[height=2in]{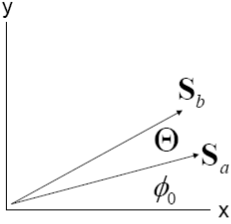}
 \caption{General spin configuration in x-y plane.}\label{fig:spinconfiguration}
\end{figure}
Also, $\nu=x,y,z$, $\mathbf{R}_a$ and $\mathbf{R}_b$ are the locations of atoms $a$ and $b$, and the origin is
at the mid-point between the two atoms. Finally the ``radial" functions $u,v$ and the one-electron potential
$V(\mathbf{r})$ are assumed to be invariant under $y\rightarrow -y$ and $z\rightarrow -z$, $V$ also being even
in $x$. We can allow hopping for any orbital to any orbital, although, again for simplicity, we assume no
intra-site transition matrix elements, and that the intra-orbital Coulomb repulsion is so large as to exclude
such inter-site hopping. We include also the intra-site inter-orbital (s-p) Coulomb repulsion $U_0$. Finally we
include the essential hopping processes $s_a\leftrightarrow p_b^\nu, s_b\leftrightarrow p_a^\nu$ caused by the
spin-orbit interaction
\begin{equation}
V_{SO}=c_o\nabla V\times \mathbf{p}\cdot\mathbf{s},\label{3}
\end{equation}
with $c_o=\hbar^2/(2m^2c^2), (\mathbf{p},\mathbf{s})$ = (momentum/$\hbar$, spin/$\hbar$).

The crucial matrix elements are, e.g.,
\begin{equation}
<p_a^\nu|V_{SO}|s_b>=c_0<\nu  v_a|\nabla V\times \mathbf{p}|u_b>\cdot<\chi_a|\mathbf{s}|\chi_b>.\label{4}
\end{equation}
I've put $v_a=v(\mathbf{r}-\mathbf{R}_a)$, etc. It is convenient to consider the spatial factor here,
\begin{equation}
m_j^\nu=<\nu v_a|(\nabla V\times\mathbf{p})_j|u_b>,
\end{equation}
where $j=x,y,z$. Using the symmetry properties of $u, v, V$ stated above, one can see readily that all the
quantities $m^\nu_i$ vanish except $m^y_z$ and $m^z_y$. The required spin factors in~(\ref{4}) are
straightforwardly found to be
\begin{eqnarray}
<\chi_a|s^z|\chi_b>&=& (1-e^{i\Theta})/4\equiv\xi_z\nonumber\\
<\chi_a|s^y|\chi_b>&=&\frac{i}{4}[e^{-i\phi_0}-e^{i(\phi_0+\Theta)}]\equiv\xi_y.\label{5}
\end{eqnarray}
We will also need
\begin{equation}
\eta\equiv<\chi_a|\chi_b>=\frac{1}{2}(1+e^{i\Theta}).\label{6}
\end{equation}

Consider first the simplest case, $\phi_0=-\Theta/2$, which implies $\xi_y=0$. Then we need only the term
$m_z^y$, which is seen to be
\begin{equation}
m^y_z=i\gamma,
\end{equation}
where
\begin{equation}
\gamma=-c_o\int d^3r\ v_a(\mathbf{r}) y \left(\frac{\partial V}{\partial x}\frac{\partial }{\partial
y}-\frac{\partial V}{\partial y}\frac{\partial }{\partial x}\right)u_b(\mathbf{r})\label{8}
\end{equation}
is real; further, there is no symmetry reason for this to vanish. Thus the remaining SO matrix element~(\ref{4})
is
\begin{equation}
M=i\gamma\xi_z.
\end{equation}

The basic symmetry of the situation allows the assumption $v_a\leftrightarrow v_b, u_a\leftrightarrow u_b$ under
$x\rightarrow -x$. It follows that $m^y_z\rightarrow -m^y_z$ under $x\rightarrow -x$, i.e. under
$a\leftrightarrow b$. The above results plus hermiticity of $\nabla V\times\mathbf{p}$ give all the relevant
matrix elements.

Because only the $p^y$ orbitals are connected to the ground state orbitals $s_a,s_b$, we can drop the other
p-states. This leaves four 1-electron states,~(\ref{1}) with $\nu=y$, and therefore six 2-electron states. We
write them conveniently in terms of $A_s^\dagger,A_p\dagger$, which respectively, create an electron in states
$s_a,p_a^y$; similarly for $B_s^\dagger$, etc.:
\begin{eqnarray}
\Phi_1&=&A_s^\dagger B_s^\dagger|0>,\ \Phi_2=A_s^\dagger B_p^\dagger|0>,\ \Phi_3=A_p^\dagger
B_s^\dagger|0>,\nonumber\\
\Phi_4&=&A_p^\dagger B_p^\dagger|0>,\ \Phi_5=A_s^\dagger A_p^\dagger|0>,\ \Phi_6=B_s^\dagger B_p^\dagger|0>.
\end{eqnarray}

Our final simplification before writing down the Hamiltonian $H$ is the Hubbard-like assumption where hopping is
a 1-electron operator, the essential contribution from the Coulomb terms being the intra-site inter-orbital
Coulomb term $U_0$. Thus, in the basis $(\Phi_1,\cdots\Phi_6)$,
\begin{equation}
H=\left( \begin{array}{cccccc} 0 & 0 & 0 & 0 &-i\gamma \xi_z^*&-i\gamma \xi_z\\
                             0 & \Delta & 0 & 0 & t'& t\\
                             0 & 0 & \Delta & 0 &-t& -t'\\
                             0 & 0 & 0 & 2\Delta & i\gamma\xi_z^* & i\gamma\xi_z\\
                             i\gamma\xi_z & t' & -t & -i\gamma\xi_z & \Delta+U_0 & 0\\
                             i\gamma\xi_z^* & t & -t' & -i\gamma\xi_z^* & 0 & \Delta+U_0\label{11}
                             \end{array}\right)
\end{equation}
The real quantities $t,t'$ are ordinary (kinetic + Coulomb) hopping terms: $t$ hops $s_a$ to $s_b$, $t'$ hops
$p_a$ to $p_b$.~\cite{other}

Let us calculate the electric dipole moment in perturbation theory, where all hoppings are small. A non-zero
value occurs in first order, surprising to us, having expected the dipole moment to come from
\emph{intra}-atomic mixing of s-like and p-like orbitals (but this occurs in higher order only). The ground
state wave function to 1st order can be picked off from the first column of $H$,~(\ref{11}), and using the
unperturbed energy of the doubly-occupied sites:
\begin{equation}
\Psi_g=\Phi_1-\frac{i\gamma}{\Delta+U_0}(\xi_z\Phi_5+\xi_z^*\Phi_6).\label{12}
\end{equation}
The dipole moment operator is $\pi=e(\mathbf{r}_1+ \mathbf{r}_2)$. To 1st order, one needs the results
\begin{equation}
<\Phi_1|\pi|\Phi_5>=-<\Phi_1|\pi|\Phi_6>^*=\hat{y}\tilde{y}<\chi_b|\chi_a>,\label{13}
\end{equation}
where
\begin{equation}
\tilde{y}=\int d^3r\ u(\mathbf{r}-\mathbf{R}_a)y^2v(\mathbf{r}-\mathbf{R}_b).
\end{equation}
Note that $\tilde{y}$ has dimensions of length. The x and z components of $\pi$ vanish by symmetry.

 Then, after a bit of arithmetic, we find
\begin{equation}
<\pi>=<\Psi_g|\pi|\Psi_g>=-\hat{y}\frac{e\gamma \tilde{y}}{U_0+\Delta}sin \Theta.\label{15}
\end{equation}
Our spins being in the x-y plane, this result is clearly consistent with~(\ref{0}).

It is instructive to consider the electron density,
\begin{equation}
n(\mathbf{r})=u_a(\mathbf{r})^2+u_b(\mathbf{r})^2-\frac{\gamma\ sin \Theta}{2(U_0+\Delta)} y\
n_{ov}(\mathbf{r}),\label{17}
\end{equation}
where the ``overlap density" is
\begin{equation}
n_{ov}(\mathbf{r})=u_a(\mathbf{r})v_b(\mathbf{r})+u_b(\mathbf{r})v_a(\mathbf{r}).\label{18}
\end{equation}
Thus the charge density responsible for the dipole moment exists mainly between the two sites. Note that the
result $\rightarrow0$ as $V_{SO}\rightarrow0$, as expected.

We now generalize to the case where $\mathbf{S}_a$ and $\mathbf{S}_b$ respectively make angles $\phi_0$ and
$\phi_0+\Theta$ with the positive x-axis, and remain in the x-y plane (see Fig.~\ref{fig:spinconfiguration}). We
will see that the dipole moment is independent of $\phi_0$.

But the non-vanishing of $\xi_y$ forces consideration of $p_a^z$ and $p_b^z$, due to~(\ref{4}) and the above
statement that $m_y^z\ne0$. Thus the number of necessary 1-electron states is increased from 4 to 6, increasing
the number of 2-electron states to 15. Nevertheless, it is again quite simple to write down the ground state to
1st order in the hopping terms, yielding the generalization of~(\ref{12}), which in turn gives the electron
density:
\begin{equation}
n(\mathbf{r})=u_a(\mathbf{r})^2+u_b(\mathbf{r})^2+2\frac{n_{ov}(\mathbf{r})}{U_0+\Delta}[\gamma
Im(\xi_z\eta^*)y+\gamma'Im(\xi_y \eta^*)z].\label{23}
\end{equation}
But from the definitions~(\ref{5}) and~(\ref{6}),
\begin{equation}
Im (\xi_y\eta^*)=0.
\end{equation}
Hence the term $\propto z$ doesn't enter, and so $\gamma'$ is irrelevant, and the result reduces to the previous
expression~(\ref{17}).

To finish this consideration of the 2-site model, we consider the case where the two spins lie in the y-z plane,
i.e. perpendicular to $\mathbf{R}$. Then the spin states are
\begin{equation}
\chi_\mu=[cos(\theta_\mu/2)\alpha+isin(\theta_\mu/2)\beta]/\sqrt2, \ \mu=a,b.
\end{equation}
In this case one can easily see that $\xi_z,\xi_y, \eta$ are all real, so that~(\ref{23}) says the dipole moment
vanishes, again consistent with~(\ref{0}).

Now apply these results to a crystal where the spins form a spiral~\cite{yoshimori, villain,kaplan2}
\begin{equation}
<\mathbf{S}_n>=1/2[\hat{x} cos(\mathbf{k}\cdot\mathbf{R}_n)+\hat{y} cos(\mathbf{k}\cdot\mathbf{R}_n)]
\end{equation}
For simplicity we can consider a cubic crystal and the propagation vector $\mathbf{k}$ along a principle cubic
direction, say x. Then $\mathbf{k}\cdot\mathbf{R}_n$ can be written $n\Theta$, where $\Theta$ is the spiral turn
angle. By choosing $\phi_0=n\Theta$, one sees from FIG.~\ref{fig:spinconfiguration} that a spiral is generated
by increasing $n$ by steps of unity. Then the induced change in electron density in the bond $j,j+1$ is
\begin{equation}
\delta n(\mathbf{r})_{j,j+1}=\frac{n_{ov}(\mathbf{r})_{j,j+1}}{2(U_0+\Delta)} \gamma y sin \Theta .
\end{equation}
That is, the dipole moment in each bond is the same (ferroelectric ordering), and the overlap density is merely
the translation of this density from one bond to the next.

This way of generating results for a crystal from those of a pair of magnetic atoms follows that
of~\cite{katsura}.

In summary, the present microscopic model yields an electric dipole moment $\mathbf{\pi}$ resulting from canted
spins, in the direction of $\mathbf{f}=\mathbf{R}\times(\mathbf{S}_a\times\mathbf{S}_b)$ with the same
dependence, $sin\Theta$, on the angle $\Theta$ between the spins. When applied to a crystal with a simple spiral
spin state, the $\mathbf{f}$ component of $\pi$ is the same for each bond, yielding a ferroelectric state, just
as in the previous closely related theory of Katsura et al~\cite{katsura}. In contrast, however, our result
$\rightarrow0$ as $V_{SO}\rightarrow0$.

We  mention a very rough estimate of the size of the effect. We assume hydrogen 1s and 2p orbitals, R=3$\AA$
(the Cr-Cr distance in CoCr$_2$O$_4$), $U_0+\Delta$=2eV, and $V$ the potential energy of an electron in the
field of 2 protons. We find the coefficient of $sin\Theta$ in~(\ref{15}) (the dipole moment for one bond), to be
$~2\times10^{-36}$C-m (Coulomb-meters). Assuming a simple cubic lattice with spiral propagation along a
principle cubic direction, and a volume per site of $(3\AA)^3$, yields a polarization of $\sim 0.1\ \mu C/m^2$,
about an order of magnitude smaller than found~\cite{yamasaki} in CoCr$_2$O$_4$. In view of the crudeness of
this estimate, it suggests that the mechanism is probably relevant to real materials.

Finally, we extended our model to the case where each site is like Cr$^{3+}$ in an octahedral field (B-site),
the three 3d-electrons being in the high-spin state $t_{2g}^3$, as in Co and Mn chromite. The p-states are from
the 4p shell. Also note that the dominant B-B exchange interaction is direct, the superexchange being
$90^{o}$~\cite{menyuk2}, suggesting that our simple model neglecting an intervening oxygen might be somewhat
realistic for the B-B pairs in these spinels. Our motivation is to check that our basic mechanism is robust in
going to a more realistic model.

The model Hilbert space is now much larger. Our basis functions are constructed as follows. The 3 $t_{2g}$
orbitals are $t^\nu(\mathbf{r})=g_\nu(\mathbf{r})u(\mathbf{r}), g_1=xy,g_2=yz, g_3=zx$. Similarly the p-orbitals
are $p^\nu(\mathbf{r})=h_\nu v(\mathbf{r}),h_1=x,h_2=y,h_3=z; u(\mathbf{r})$ and $v(\mathbf{r})$ are invariant
under inversion, $\mathbf{r}\rightarrow-\mathbf{r}$. The full 1-electron basis states are
\begin{eqnarray}
T_{\stackrel{a}{b}}^\nu&=&t_{\stackrel{a}{b}}^\nu(\mathbf{r})\chi_{\stackrel{a}{b}}\nonumber\\
P_{\stackrel{a}{b}}^\nu&=&p_{\stackrel{a}{b}}^\nu(\mathbf{r})\chi_{\stackrel{a}{b}};
\end{eqnarray}
$t_a^\nu(\mathbf{r})=t^\nu(\mathbf{r}-\mathbf{R}_a)$, etc. We again simplify somewhat by choosing the spin
states $\chi_{\stackrel{a}{b}}$ as in~(\ref{2}) with $\phi_0=-\Theta/2$. Let $A_t^{\nu\dagger}, A_p^\nu$ create
$T_a^\nu,P_a^\nu$ respectively, and similarly for $B_t^{\nu\dagger},B_p^{\nu\dagger}$. Write these as
$C_\gamma^\dagger, \gamma=1,\cdots,12$, with $\gamma=1,\cdots,6$ corresponding to the T-states,
$\gamma=7,\cdots,12$ to the P-states.

Then the spin-orbit interaction is conveniently written
\begin{equation}
V_{SO}\equiv c_0\sum_{i=1}^6\nabla_iV\times\mathbf{p}_i\cdot\mathbf{s}_i
=\sum_{\gamma,\gamma^\prime}<\gamma|v^{so}(\mathbf{r},s)|\gamma^\prime>C_\gamma^\dagger C_{\gamma^\prime},
\end{equation}
where $v^{so}(\mathbf{r},s)=c_0\nabla V\times\mathbf{p}\cdot\mathbf{s}$. Then the 6-electron basis states
(single determinants) are
\begin{eqnarray}
\Phi_1&=&\Pi_1^6C_\gamma^\dagger|0>\equiv|0),\nonumber\\
\Phi_{\gamma^{\prime}\gamma}&=&C^\dagger_{\gamma^{\prime}} C_\gamma|0),\ \gamma\le6,\ \gamma^{\prime}>6.
\end{eqnarray}
For the hole vacuum $|0)$, the $C_\gamma^\dagger$ are ordered $C_1^\dagger C_2^\dagger \cdots C_6^\dagger$. The
ground state to 1st order in the overlap is then
\begin{equation}
\Psi_g=\Phi_1-\sum_{\gamma\le6,\gamma^{\prime}>6}
\frac{\Phi_{\gamma^{\prime}\gamma}<\gamma^{\prime}|v^{so}|\gamma>}{E_{\gamma^{\prime}\gamma}^0-E_1^0},
\end{equation}
where the $E^0$ are the unperturbed energies. The terms that contribute to the electric dipole moment are the
interatomic elements, e.g.
\begin{eqnarray}
<P_a^\nu|v^{so}|T_b^\mu>&=&c_0<p_a^\nu|\nabla V\times\mathbf{p}|t_b^\mu>\cdot <\chi_a|\mathbf{s}|\chi_b>\nonumber\\
&\equiv& i\mathbf{o}_{\nu\mu} \cdot <\chi_a|\mathbf{s}|\chi_b>.
\end{eqnarray}
Most of the matrix elements of $\mathbf{o}$ vanish by symmetry; noting that the y-component of the spin matrix
element $\xi_y$ vanishes (by the choice of $\phi_0$, as before), we need only the components $o^i, i =x,z$. The
only ones of these matrix elements that survive are
\begin{eqnarray}
io_{11^{\prime}}^z&=&c_0<p_a^1|(\nabla V\times\mathbf{p})_z|t_b^1>\nonumber\\
io_{32^{\prime}}^z&=&c_0<p_a^3|(\nabla V\times\mathbf{p})_z|t_b^2>
\end{eqnarray}
plus 3 terms for $o^x$. But in calculating $\pi$, these matrix elements get multiplied by corresponding
elements of the position or displacement operator as in
$\mathbf{r}_{\gamma\gamma^{\prime}}v^{so}_{\gamma^{\prime}\gamma}$, and many of the position matrix elements
vanish by symmetry. It turns out that only the $o^z$ terms contribute, and the final result for the dipole
moment is
\begin{equation}
<\pi>=-\frac{e\hat{y}}{U_0+\Delta}sin\Theta(\gamma_1\tilde{y}_1+\gamma_2\tilde{y}_2),
\end{equation}
where
\begin{eqnarray}
\gamma_1&=&<p_a^1|o^z|t_b^1>,\ \gamma_2=<p_a^3|o^z|t_b^2>\nonumber\\
\tilde{y}_1&=&\int d^3 r p_a^1 y t_b^1=(1/2)\int d^3r (x^2-R^2/4)n_{ov}\nonumber\\
\tilde{y}_2&=&\int d^3 r p_a^3 y t_b^2=(1/2)\int d^3r y^2z^2n_{ov}.
\end{eqnarray}
Thus the result is similar to the previous one~(\ref{15}). The terms, $\tilde{y}_i\gamma_i, i=1,2$ do not vanish
by symmetry. Thus we have confirmed that the basic mechanism, involving inter-atomic SO coupling, gives the
interesting ME effect in this rather realistic model.

One of us (T.A.K) thanks C. Piermarocchi, J. Bass, and M. Dykman for helpful discussions.

\end{document}